\documentclass{sigchi}

\usepackage{balance}       
\usepackage{graphics}      
\usepackage[T1]{fontenc}   
\usepackage{txfonts}
\usepackage{mathptmx}
\usepackage[pdflang={en-US},pdftex]{hyperref}
\usepackage{color}
\usepackage{booktabs}
\usepackage{textcomp}
\usepackage{multirow}

\usepackage{microtype}        
\usepackage{ccicons}          

\usepackage{todonotes}
\usepackage{subfig}

\def\plaintitle{Kid on The Phone! Toward Automatic Detection of Children on Mobile Devices}

\def\emptyauthor{}
\def\plainkeywords{Children; Adults; Age Group; Behavior; Mobile Devices; Parental Control; Online Safety}

\makeatletter
\def\url@leostyle{%
  \@ifundefined{selectfont}{
    \def\UrlFont{\sf}
  }{
    \def\UrlFont{\small\bf\ttfamily}
  }}
\makeatother
\def\pprw{8.5in}
\def\pprh{11in}

\definecolor{linkColor}{RGB}{6,125,233}
\hypersetup{%
  pdftitle={\plaintitle},
  pdfauthor={\emptyauthor},
  pdfkeywords={\plainkeywords},
  pdfdisplaydoctitle=true, 
  bookmarksnumbered,
  pdfstartview={FitH},
  colorlinks,
  citecolor=black,
  filecolor=black,
  linkcolor=black,
  urlcolor=linkColor,
  breaklinks=true,
  hypertexnames=false
}

\begin{document}

\title{\plaintitle}

\numberofauthors{3}
\author{%
 \alignauthor{Toan Nguyen\\
  \affaddr{New York University}\\
    \email{toan.v.nguyen@nyu.edu}}\\
  \alignauthor{Aditi Roy\\
    \affaddr{New York University}\\
    \email{ar3824@nyu.edu}}\\
 \alignauthor{Nasir Memon\\
    \affaddr{New York University}\\
    \email{memon@nyu.edu}}\\
}

\maketitle

\begin{abstract}
  Studies have shown that children can be exposed to smart devices at a very early age. This has important implications on research in children-computer interaction, children online safety and early education. Many systems have been built based on such research. In this work, we present multiple techniques to automatically detect the presence of a child on a smart device, which could be used as the first step on such systems. Our methods distinguish children from adults based on behavioral differences while operating a touch-enabled modern computing device. Behavioral differences are extracted from data recorded by the touchscreen and built-in sensors. To evaluate the effectiveness of the proposed methods, a new data set has been created from 50 children and adults who interacted with off-the-shelf applications on smart phones. Results show that it is possible to achieve 99\% accuracy and less than 0.5\% error rate after 8 consecutive touch gestures using only touch information or 5 seconds of sensor reading. If information is used from multiple sensors, then only after 3 gestures, similar performance could be achieved.
\end{abstract}

\keywords{\plainkeywords}

\section{Introduction}
\label{sec:kidsonthephone_introduction}
The use of mobile devices and the internet by children has increased rapidly in the past decade, approaching saturation by middle childhood in developed countries. Research has been published, exploring both positive and negative consequences of this increase~\cite{wisniewski2015resilience}. Important research being done includes topics such as children online safety~\cite{hartikainen2016should, wisniewski2015resilience}, parental control~\cite{zaman2016parental, nouwen2015value}, benefits of early education using mobile technology, inclusion of children in system and user interface (UI) design~\cite{woodward2016characterizing, vatavu2015touch}. Systems and applications have also been built based on these research efforts.

However,  the above systems often assume knowledge of the presence of a child on the device. In other words, they assume that a child is using the device and the system will operate in a child-friendly mode. 
In addition, most parental control apps require parents to activate a restriction feature before giving the phone to the child. If they forget, the child may not be restricted at all. Having the ability to detect whether a child is using the device on the fly is desired, as it can significantly improve the reliability and user experience of these systems. 

In this paper, we present techniques to automatically detect whether a child is on a mobile device. This technology, if enabled, can be used as the first step for  other  systems. For example, when a system detects that a child is using the device, it could switch to a more child-friendly UI or turn on parental control. We propose three new approaches to model the behavioral difference in the use of mobile devices between children and adults. A new data set was acquired which includes touch and sensor data of 25 children (age 3--12) and 25 adults (age 24--66) captured while they were operating a mobile device. We also conducted a comprehensive set of experiments to find the best classifiers as well as evaluate the performance of each approach. The results show that 99\% detection accuracy can be achieved with a low error rate of less than 0.5\% for all three approaches. We complete our study with a discussion on the strengths and weaknesses of the different approaches we have studied. 

The rest of the paper is organized as follows. We survey related work in the next section and then present an overview and motivation of our research. Details of three different children detection approaches are presented in subsequent sections. A data collection section is presented next followed by evaluation results of proposed methods. We then  summarize our work as well as its limitations and discuss broader impacts and provide some avenues for future work in the last section. 

\section{Related work}
\label{sec:kidsonthephone_related_work}
Modern mobile devices provide a rich set of sources for extracting user behavioral data. The ubiquitous touchscreen and a variety of sensors such as audio sensors, light sensors, accelerometers, gyroscope, and magnetometers can be leveraged  for capturing user behavior data \cite{SMI12b, de2012touch, 6786375, HLM12, HMS12, van2014finger, nguyen2017draw, lee2017implicit, nguyen2017smartwatches, papadopoulos2017illusionpin, NGUYEN2018174}. Most of such work, however, has focused on user authentication as described below.

\subsection{User authentication from touch and sensor data}
Existing approaches for behavioral user authentication on mobile devices can be broadly divided into two types based on the data sources used, namely, touch-based and sensor-based methods.

\subsubsection{Touch-based user authentication}
With the increased usage of mobile devices and the known limitations of passwords, user touch characteristics have emerged as a promising behavioral biometric that can be measured~\cite{SilentSense13, Chan14, van2014finger, jain2015exploring, nguyen2017draw, NGUYEN2018174} for the purposes of authentication. Since Kim et al. \cite{kim2010multi} first explored the possibility of users being authenticated by their touch-interaction behavior, there has been a growing body of work in this direction \cite{SMI12b, de2012touch, 6786375, van2014finger, nguyen2017draw, NGUYEN2018174}. It has been observed that touch trails made on the screen of a mobile device are quite distinctive among users in terms of the spatio-temporal patterns made by their fingers, as well as the area and pressure of the touch. A wide range of techniques have been proposed that take different approaches to acquire user input and use different verification methods. Users, for example,  may need to perform a certain specified set of gestures to authenticate~\cite{de2012touch, shahzad2013secure, van2014finger, nguyen2017draw} or may freely draw on the screen~\cite{sherman2014user}. User verification can be done with widely used pattern recognition algorithms like Dynamic Time Warping~\cite{de2012touch, van2014finger, nguyen2017draw}, or more complex machine learning frameworks~\cite{shahzad2013secure, song2017multi}. In addition to these point of entry authentication techniques, continuous authentication has also emerged as a promising approach to alleviate security problems that stem from poor authentication technology \cite{roy2014hmm, ray2015, patel2016continuous}. In this approach, users are periodically re-authenticated during a session based on their touch characteristics~\cite{Frank13, meng2013touch, zhao2014mobile, feng2014tips, antal2015information}.

\subsubsection{Sensor-based user authentication}
A large body of user authentication research has focused on extracting behavioral patterns from mobile sensors~\cite{Senguard11, jain2015exploring, ray2015, lee2017implicit}. The general idea is that user movements can be inferred from sensor data recorded while they are interacting with a mobile device and this movement pattern is distinctive enough to authenticate users. The authentication performance reported  varies greatly, from as low as 75\% accuracy~\cite{zhu2013sensec} to as high as over 99\% accuracy~\cite{juefei2012gait}. 

However, all of the above mentioned studies focus on capturing the distinctiveness of an individual user for the purpose of authentication. In contrast, our work seeks to find common characteristics in touch behavior of an age group. This is a more difficult task since the behavior variation between different individuals within the same group may be large. Moreover, the overlap in behavior between different groups may pose a significant challenge to overcome. 

\subsection{Children age group detection}
In contrast to the research on behavior biometric based user authentication in mobile devices, studies on children age group detection is sparse. Approaches have been proposed using speech and facial features~\cite{meinedo2010age, li2013automatic, qawaqneh2017age}. However, these approaches not only need explicit user permission to use the microphone and the camera, but their performance is also significantly affected by external conditions. For example, users may not speak a word during a session. Also, it may not be easy to capture an image with sufficient quality for detection (i.e., images may have only a part of user's face or may be blurred due to shaking of user's hand). Therefore, an implicit and less intrusive detection system is needed that runs in the background. 

Recently, Vatavu et. al~\cite{vatavu2015child} proposed an approach to detect small children (3--6) from tap gesture analysis on mobile devices. Hernandez et. al~\cite{hernandez2017detecting} used the Sigma-Lognormal model to extract neuromotor characteristics obtained from drag and drop gestures for age group classification. Both studies used the same data set and consequently had the same limitation: requiring fixed type of gestures and only including small children (age 3--6) in their work. This is a significant drawback because these techniques may not work for older children whose behavior is more similar to adults.

Another work by Li et. al~\cite{li2018icare} extended the age group of the children subjects to 3--11. They used taps and swipes for extracting user behavior. However, only limited experiments were carried out. For example, in their evaluation, they did not have enough data to do classification of older children based on tap data. While the three above-mentioned works focused only on touch behavior, Davarci et. al~\cite{davarci2017age} proposed children age group detection based on accelerometer reponses when users tap on the screen. They measured the difference in accelerometer readings between different users when they tap on the screen and used it for detection of children subjects. They did not model user movements when holding the device. 



\subsection{This work}
\label{sec:kidsonthephone_general_idea_and_overview}
In this work, we assume that a child is sharing a mobile device with his or her parent. The objective of this work is to determine who is operating the device (a child or an adult) by analyzing the behavior of the user. If it is detected that the user is a child, the desired action can be taken like turning on parental control or switching the UI. The system should be unobtrusive, quick, accurate and user-agnostic while being passive and continuous. The general idea is that data recorded from several touch gestures and/or sensor readings is distinctive enough to distinguish between children and adults. We divide solutions into touch-based, sensor-based and a combination of touch--sensor approaches. 

In the touch-based approach, the touchscreen of the device captures each touch event and generates corresponding touch data sequence. Behavioral features are extracted from this data to model a user's behavior. The intuition is that children have smaller fingers which results in smaller touch area on the screen. In addition, they tend to swipe faster than adults do, which produces shorter and less curvy swipes. Our experimental results confirm these observations.

In the sensor-based approach, several sensors available on modern devices like accelerometers and gyroscopes are used to capture users' movements. Our hypothesis is that children, with smaller hands, will tend to be more shaky than adults when holding the device. Also, children are more active, so they will move the phone more often when handling it, which can be inferred from sensor data.

In the touch-sensor combination approach, we combine features extracted from a touch gesture, and sensor data recorded during this gesture, for complete user behaviour model. Sensor data is captured when the user touches the screen instead of continuously recording as in the sensor-based approach. This will help reduce battery consumption and computational overhead and still achieve high detection performance of both touch and sensor-based classifiers.

Our goal is to explore a generic, user-independent model for children detection. This means the model is built on aggregated data of multiple users in the same age group. 




The proposed child detection framework works in two steps: training and detection. During training, a behavioral model is created based on the behavioral features from a training data set. 
These features can be touch, sensor or a combination of both, depending on the approach. The model is stored on the device. Later, when a user picks up the device and starts using it, various data (touch and sensors) will be recorded and fed into a feature extraction process, which then inputs the extracted feature vector to the stored model for a prediction. If the model predicts the user to be a child, then a desired action will be activated on the device based on the settings of the application. Note that we would like to build a generic model that can be loaded into any phone or mobile applications. It is not tailored to each phone or each user.




\section{Touch-based Approach for Children Detection}
\label{sec:kidsonthephone_touch-based-approach}

In this section, we explore feature selection and the classification framework of the touch-based children detection approach. 

A touch-screen based smart device is typically operated by touch commands such as tap, stroke (includes slide, fling, scroll), pinch and free stroke handwriting. However, pinch, a two-finger  gesture  to  do zooming, is used less than 5\% of the time and perhaps even less by children~\cite{Frank13, roy2014hmm}. Similarly, though handwriting is an alternative input method to enter characters, it is not used very often. Tap and stroke are the basic and most frequently used gestures. So, similar to previous work~\cite{Frank13, roy2014hmm}, we used only these two gestures in our study. We  build separate models for each type of gesture.


\subsection{Feature extraction for tap gestures}
\label{subsec:kidsonthephone_tap_feature_extraction}
A raw tap gesture includes several touch events (i.e., touch down, touch up). To represent a tap, we average values of touch events into a single tap. So, each tap has the following values: x,y-coordinates, pressure, size of touch, duration of tap (calculated by subtracting timestamp of the first touch event from timestamp of the last touch event in the tap).
We extract (\emph{pressure, size, and duration}) as the feature vector to represent a tap in our evaluation. We discard x,y coordinates, which are highly personal features~\cite{Frank13, roy2014hmm, xu2014towards}.

\subsection{Features extraction for stroke gestures}
\label{subsec:kidsonthephone_touch_features_extraction}
A stroke is a sequence of touch events starting with a touch down and finishing with a touch up event. We extract 22 features from a stroke (see Table~\ref{tbl:kidsonthephone_stroke_features}). Most of the features are self-explanatory.The Largest Deviation Point (LDP) is a point in the stroke that is farthest from the straight line drawn from the start and end touch points of the stroke. This term was first used in~\cite{xu2014towards}. Some other notable features are \emph{average\_size} and \emph{std\_size} which are average and standard deviation of all size values in the stroke. Velocity is calculated by dividing the distance between two consecutive touch points in the stroke over the duration between them (calculated by subtracting two timestamps of two touch points). The distance between two touch points (\emph{$x_i, y_i$}) and (\emph{$x_j, y_j$}) is defined as $\sqrt{(x_j-x_i)^2 + (y_j-y_i)^2}$. 

To rank features, we trained a Random Forest with stroke data from a new data set consisting of children and adults touch data (described in data collection section below) and configured it to return feature importance scores using \emph{scikit-learn}~\footnote{http://scikit-learn.org}, a popular Python package for machine learning. It's notable that two of the top-3 features are extracted from the size or area of the touch. This shows that size is an important factor. The top ranked feature is \emph{straight\_to\_trajectory\_length\_ratio}, which is calculated as the ratio between straight line between the start and stop touch point of a stroke, and the trajectory which connects all touch points in the stroke. This feature indicates how curvy a stroke is. We expect that children tend to perform short and fast slides/scrolls, resulting in less curvy strokes. 

\begin{table}[!t]
\centering
\caption{Stroke features ranked by importance scores.}
\label{tbl:kidsonthephone_stroke_features}
\begin{tabular}{lll}
Rank & Feature name                            & Importance score \\ \hline
1    & straight\_to\_trajectory\_length\_ratio & 0.076282         \\
2    & average\_size                           & 0.073745         \\
3    & std\_size                               & 0.057999         \\
4    & start\_p                                & 0.052019         \\
5    & start\_to\_LDP\_length                  & 0.050034         \\
6    & average\_velocity                       & 0.049375         \\
7    & std\_velocity                           & 0.049014         \\
8    & start\_to\_LDP\_duration                & 0.047101         \\
9    & average\_pressure                       & 0.046342         \\
10   & LDP\_to\_stop\_length                   & 0.044749         \\
11   & trajectory\_length                      & 0.043983         \\
12   & average\_distance                       & 0.043941         \\
13   & straight\_length                        & 0.043939         \\
14   & LDP\_velocity                           & 0.043720         \\
15   & std\_distance                           & 0.042575         \\
16   & std\_pressure                           & 0.041388         \\
17   & LDP\_to\_stop\_duration                 & 0.040684         \\
18   & LDP\_p                                  & 0.039430         \\
19   & stop\_p                                 & 0.038895         \\
20   & LDP\_s                                  & 0.038578         \\
21   & start\_s                                & 0.022431         \\
22   & stop\_s                                 & 0.013776        
\end{tabular}
\end{table}

\begin{figure*}[!t]
    \centering
    \subfloat[Ranking of all sensor features]{{\includegraphics[width=\textwidth]{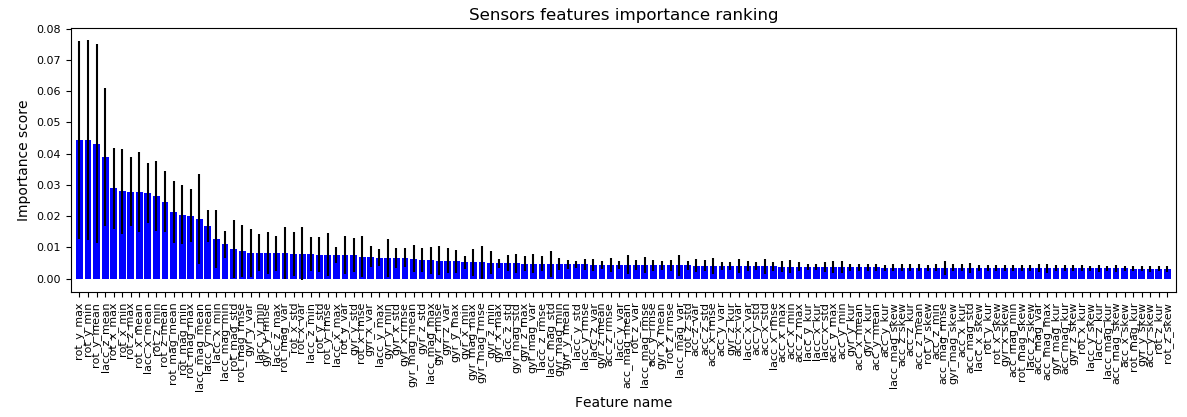} }}%
    \hfill
    \subfloat[Top 20 sensor features]{{\includegraphics[width=0.4\textwidth]{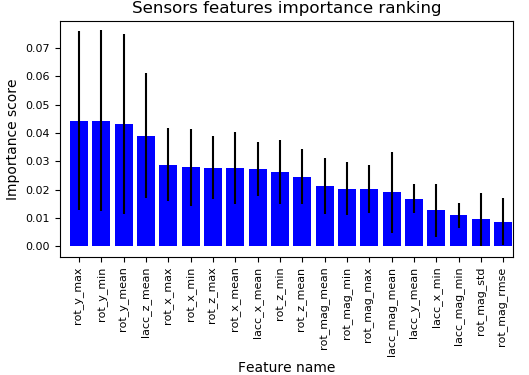} }}%
    \caption{Sensor features ranking and selection.}%
    \label{fig:kidsonthephone_sensor_features_ranking}%
\end{figure*}

\subsection{Classification}
Since the touch behavior differs significantly for each individual, we expect a large variance in touch features both in case of children and adults. Hence, we need a strong classifier to do the classification. Support Vector Machines (SVM)~\cite{cortes1995support} and Random Forests (RF)~\cite{breiman2001random} are great candidates. SVM has been shown to perform well on touch-based authentication~\cite{Frank13, xu2014towards, shahzad2013secure}. In our binary classification problem (child/adult), an SVM builds a hyperplane to separate the feature space such that it maximizes the margin between the two classes. A Random Forest (RF) is another state-of-the-art robust classifier, which usually produces similar or better results than SVM. 
The forest is used to classify a feature vector by inputting the vector down every tree in the forest. Each tree outputs a classification label and the forest chooses the class with the most ``votes'' among the trees as the final prediction. RF is also easier and faster to train than SVM.


Similar to touch-based authentication~\cite{Frank13, xu2014towards, roy2014hmm}, we can not expect that a single gesture (tap or stroke) would be enough to distinguish a child from adults. To increase the robustness of the detection method, multiple consecutive gestures are used for the final decision. To achieve this, we configure our classifiers to return a score -- a probability of each class (child or adult) for each gesture. We then average scores of multiple gestures and return the final decision based on a threshold. The threshold can vary depending on  the security sensitivity of the application being used.

\section{Sensor-based approach for children detection}
\label{sec:kidsonthephone_sensors-analysis}
In this section, we propose a child detection approach which uses sensor data for classification. The idea is that the way children hold and move the device is different from adults. Their hands are smaller and weaker, so they shake the device more. Children are typically more active than adults, and presumably move the device around more often while using it. We describe how these differences in behavior can be extracted from multiple motion sensors readily available on mobile devices. 

\subsection{Sensor selection}
Modern mobile devices are usually shipped with a broad range of built-in sensors including environmental, position and motion sensors. 
Motion sensors measure acceleration and rotational forces of the device along three axes using accelerometers, gyroscopes, linear acceleration and rotational vector sensors. Since we are interested in measuring the movement behavior of the user active on the device, we opt to only use motion sensors. In addition, including other sensors that may be influenced by the environment may introduce noise unrelated to the user's behavior~\cite{lee2017implicit}. We present a brief overview of the four motion sensors we choose to use for feature extraction.

Accelerometer and gyroscope have been used extensively in sensor-based work~\cite{lee2017implicit, ray2015, kwapisz2011activity, Senguard11}. Accelerometer measures the velocity change rate over time of the device along three axes. Gyroscope measures the device's rate of rotation around each of the three physical axes. Linear acceleration can be a software or hardware sensor which measures the acceleration force that is applied to the device on all three physical axes, excluding the force of gravity. It is usually calculated from the accelerometer with the help of gyroscopes to remove gravity force. Rotation sensor measures the direction change of the device along three axes.

\begin{figure*}[!t]
    \centering
    \subfloat[Top-50 of tap-sensor combined features]{{\includegraphics[height=2in, width=\textwidth]{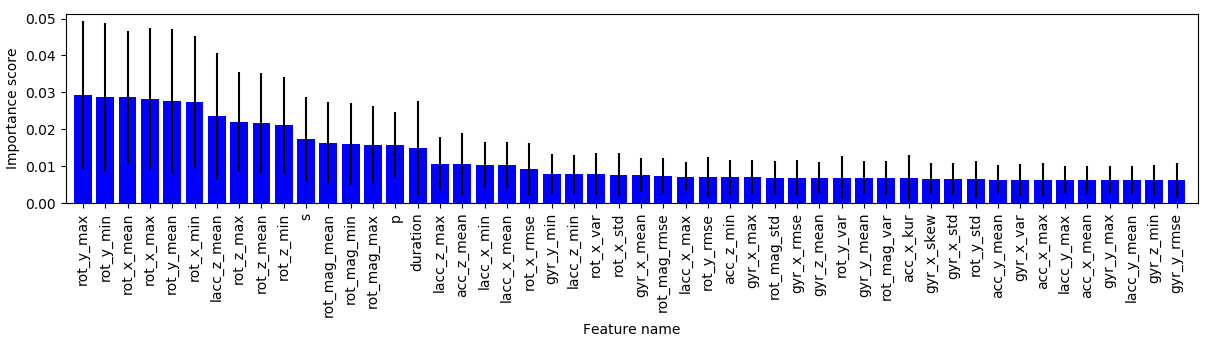} }}%
    \hfill
    \subfloat[Top-50 of stroke-sensor combined features]{{\includegraphics[height=2in, width=\textwidth]{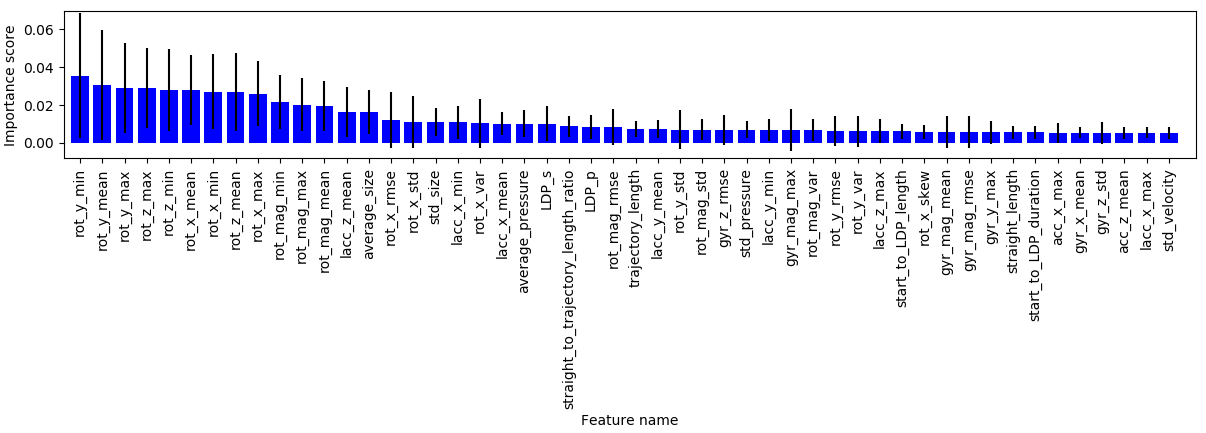} }}%
    \caption{Touch--sensor combination approach: Top 50 ranked features of the combined tap--sensor and stroke--sensor features.}%
    \label{fig:kidsonthephone_top50_combined_features}%
\end{figure*}

\subsection{Feature extraction and selection}
\label{subsec:kidsonthephone_sensors_preprocessing}
The raw data of each sensor on mobile device is a timeseries  with the following values: readings of three axes (x, y, z) and timestamp \emph{t} in UTC format. We compute and add magnitude values to the data stream. The magnitude of a sample ($x,y,z,t$) from a sensor is computed as $mag=\sqrt{x^2+y^2+z^2}$. 
The processed sensor data is then used for feature extraction.

Typically, using raw sensor data streams as input for classifiers does not provide the desired performance as individual sensor values hardly capture the essence of a user's behavior. Common practice is to segment the sensor stream into multiple time windows and extract a feature vector from each window~\cite{lee2017implicit, ray2015, kwapisz2011activity, Senguard11}. These feature vectors would then be fed into a classifier for training and testing. We use the same approach and compute following eight features from each time window of an axis of each sensor: \emph{mean, standard deviation (std), variance (var), min, max, root-mean-square deviation, skewness} and \emph{kurtosis}. Most features are self explanatory. Here, 
\emph{root-mean-square deviation} is computed as:
\begin{equation}
    \sqrt{ \frac{ \sum_1^n (s_t- \widehat{s}_t)^2 }{n} }
\end{equation}
where $s_t$ is the value of a sensor axis in a time window that contains $n$ samples, $\widehat{s}_t$ is the mean value. \emph{Skewness} and \emph{kurtosis} measure statistical characteristics of the sensor time window, whether it is symmetric or heavy-tailed relative to a normal distribution. They are computed as:
\begin{equation}
    Skewness = \frac{\sum_1^n (s_t - \widehat{s}_t)^3}{n \times std^3}
\end{equation}
\begin{equation}
    Kurtosis = \frac{\sum_1^n (s_t - \widehat{s}_t)^4}{n \times std^4}
\end{equation}

In total, we extract 8 features/axis $\times$ 4 axes (x, y, z, magnitude) $\times$ 4 sensors = 128 features from a time window.

Feature selection can then be done as follows. Features of all sensor segments in the training data are extracted using a window size of \emph{n} seconds. A Random Forest model is then trained on this feature set and is configured to return the importance score of each feature. 
The ranking of all features we obtained is shown in Figure~\ref{fig:kidsonthephone_sensor_features_ranking} (a) ( window size $\emph{n}=1$ second). Here, each feature is named in the format \emph{\{sensor name abbreviation\}\_\{axis\}\_\{feature name\}}. For example, \emph{lacc\_x\_mean} indicates the mean value of x axis of the linear acceleration sensor. To improve classification performance and speed up the training and testing process, we picked the top 20 features as depicted in Figure~\ref{fig:kidsonthephone_sensor_features_ranking}(b). It is interesting to note that the top 20 features are all extracted from rotation and linear acceleration sensors. This is because these two sensors are used to measure subtle movement of the device without unrelated factors like gravity and are computed from raw data of accelerometers and gyroscopes. Thus, they are more effective in capturing user behavior and provide better features. Note that we can vary window size \emph{n} to find the one that gives the best classification performance. The feature ranking results presented here are based on window size of $\emph{n}=1$ second. However, we obtain similar results with different window sizes. In the evaluation section, we will show the results of this procedure. 

\subsection{Classification}
\label{subsec:kidsonthephone_sensor_based_classification_framework}
Similar to touch-based approach, we want to build a generic model based on sensor data for the child detection task. 
We continue to choose SVM and RF as our classifiers as they have shown great performance on sensor based methods. In addition, we combine predictions of multiple time windows to achieve a more reliable final decision. Therefore, this technique can also be used for continuous detection of children over the period of a session.

\section{Multi-sensor touch-based approach for children detection}
\label{sec:kidsonthephone_combination-analysis}

In this section, we present an approach that combines touch and sensor features for children detection. The idea is to only record sensors whenever a user touches the screen instead of constantly recording sensors. This would help reduce battery consumption and computational overhead of the detection system and improve detection accuracy. The feature extraction module will extract features from both data sources, touch and sensors.

\begin{figure*}[t]
    \centering
    \subfloat[AUCs]{{\includegraphics[width=0.4\paperwidth]{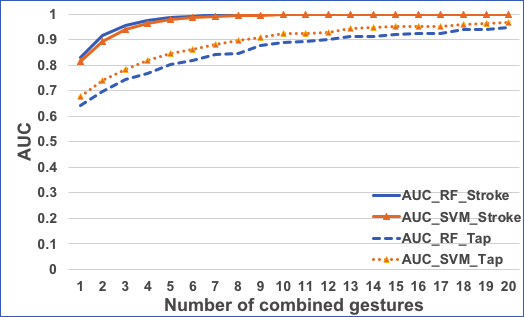} }}%
    \subfloat[EERs]{{\includegraphics[width=0.4\paperwidth]{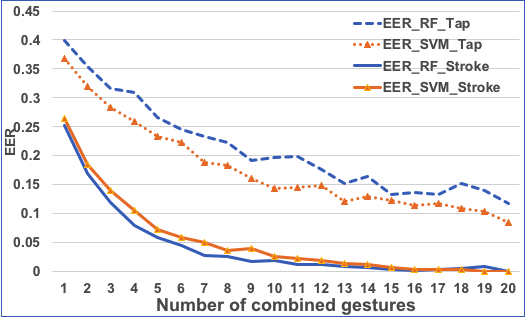} }}%
    \caption{Touch-based approach: Performances of SVM and RF when using tap and stroke features for children/adults classification. Using strokes resulted in higher AUC and lower EER. In addition, bundling multiple gestures for a classification decision significantly boosted AUC and lowered EER.}%
    \label{fig:kidsonthephone_tap_stroke_classification_power}%
\end{figure*}

\subsection{Feature extraction and combination}
\label{subsec:kidsonthephone_feature_extraction_combination_approach}
In this approach, two sets of features are extracted from a touch gesture and its corresponding sensors data. First, touch features 
are extracted from touch sequences and are similar to those in the touch-based approach. Second, sensor features, which are similar to those in the sensor-based approach, are extracted from a sensor window equals to the duration of the touch gesture. The two feature sets are then combined into one vector to represent a gesture.

Similar to previous approaches, we train a Random Forest model on a data set (described in the following section) and configure it to return importance scores of combined \emph{tap$+$sensors} and \emph{stroke$+$sensors} features. The top 50 features of each type are listed in Figure~\ref{fig:kidsonthephone_top50_combined_features}. \emph{Size of finger} and \emph{pressure} of tap features are both in the top 15 combined \emph{tap$+$sensors} features. 10 stroke features are among the top 50 combined \emph{stroke$+$sensors} features. These results show that touch features are still important in modelling user behavior.

For classifier selection, we continue to choose SVM and RF.
In the subsequent sections, we present our study to evaluate our proposed models for children detection.

\section{Data acquisition}
\label{sec:kidsonthephone_data_acquisition}
We conducted a study on Android phones and collected a new data set, which included children and adults (mostly parents of children subjects). The study was approved by our institution's research ethics board. Children subjects were given verbal assents and their parents completed written parental permission forms. If parents also participated, they had to sign independent consent forms.

\subsection{Study procedure}
\label{subsec:kidsonthephone_procedure}
The study included three sessions. In the introductory session, subjects were given a tutorial on what they were expected to do in the study. Subjects were also given a Nexus 6 phone to get familiar with. In the main session, after an experimenter turned on the logging app and put it in the background mode, the device was given to subjects to do the following tasks. First, they opened the photo app, opened two images and identified differences between them. This task was repeated twice with two different image pairs. The purpose of this task was to capture the behavior when subjects browse photos on their phone. To spot the differences between two images, they needed to swipe to switch between the two images (they could not see both images at the same time). Whenever they spotted a difference, they tapped on its position on an image.

Second, the subjects played a game in which they performed simple gestures like tap, scroll, swipe up/down/left/right to make cartoon objects displayed on the screen disappear. Third, they were asked to open the YouTube app and browse/search for videos of their choice and watch them for a few seconds. This task was repeated several times for different videos. We also planted some seed videos (mostly nursery rhymes) for smaller subjects who were not able to read or search to browse and click on and watch. All subjects were asked to do the experiment while sitting and holding the phone in their hands. In the last session, an exit survey was given to the adult subjects who had children to answer questions about mobile device usage of their children and their opinion of our study. Overall, the study lasted about 45 to 60 minutes. Each subject was compensated with a \$10 gift card.

\subsection{Apparatus}
\label{subsec:kidsonthephone_apparatus}
We developed a logging app in Java to collect touch data on Android devices. The app had two services running in the background to capture data. A touch-logger service that used Android's \textit{getevent} API to capture all touch events reported by the OS. This allowed us to capture any gesture on any application. Each touch event data included an event code (finger up/down/move), coordinates (x-y), pressure and size of the touch and a timestamp in UTC format. A gesture recognizer was built in the logging app to recognize subject gestures. The app recorded each gesture along with all of its touch points, app id (which app the subject was on when this data was recorded), device orientation and data label (child or adult, set by the experimenter before recording). 

A second service, sensor-logger, was used to capture readings of seven sensors on the device when the subjects are using it. These sensors were accelerometer, gyroscope, magnetometer, gravity, linear acceleration, rotation and orientation. The first four are physical sensors available on most modern mobile devices. The latter three are logical sensors whose readings are calculated from the physical sensors and provided via Android API to the developers for rapid app development. Each sensor reading was a tuple of four values: x, y, z value and a timestamp in UTC format. We set the recording rate for all sensors at 100 Hz.

After a subject finished the study, the experimenter turned off the app (which stopped the two services), extracted the data out of the phone and saved it on a laptop before resetting for the next subject.

\subsection{Subjects}
\label{subsec:kidsonthephone_subjects}
50 subjects (25 children and 25 adults) were recruited through  advertisement on social networks. Subjects who were under 13 years of age were considered child subjects in our study. We followed the definition of children in the Children's Online Privacy Protection Rule (``COPPA'')~\cite{coppa}.
Adults (average age=36, min=24, max=66) were mostly parents of the child subjects. Child subjects were divided into two groups: young children (19 subjects, average age=5.57, min=3.5, max=8) and older children (6 subjects, average age=10, min=9, max=12).

\subsection{Data summary}
\label{subsec:kidsonthephone_data_summary}
A filtering step was applied to each subject's data to remove data introduced by the experimenter (any data from the beginning and at the end of the data files, which had app id of the logging app). We then removed values in the data that have been shown to be highly contextual or personal behavioral features~\cite{Frank13, xu2014towards}. These values included x,y-coordinates of touch data, device orientation, and stroke direction. The reason for this removal was that we want to build generic and robust models for children and adults, thus, we opted to use generic data and less context dependent features which would be influenced by the underlying application being used.

As mentioned before, we used only tap and stroke gestures similar to previous studies~\cite{Frank13, roy2014hmm}. In total, 14383 gestures (5278 taps, 9105 strokes) and over 1.1GB of raw sensor data was collected. 

\section{Experimental results}
\label{sec:kidsonthephone_evaluation_results}

In this section, we report evaluation results of the three proposed models for children/adults classification. We report the performance of our classifiers using popular metrics: AUC -- Area Under the Receiver Operating Characteristic (ROC) curve and Equal Error Rate (EER). ROC curve is the most common way to visualize the performance of a binary classifier. It is created by plotting True Positive Rate (TPR) against False Positive Rate (FPR) at various threshold values. AUC is arguably the best way to measure a classifier's accuracy. Its value ranges from 0 to 1, where 1 is a perfect accuracy score. On the other hand, EER is considered the best measure for describing a classifier's error rate. EER is the point where False Positive Rate and False Negative Rate (FNR) are equal. EER balances between usability and security. The lower the EER, the higher the accuracy and consequently, the better the classifier is. In our evaluations, we label children samples as positives and adult samples as negatives. So FPR indicates the percentage of adult samples being misclassified as children and FNR is the percentage of children samples misclassified as adult.

\subsection{Touch-based Classifier}
\label{subsec:kidsonthephone_touch_results}

We used \emph{sklearn} to implement and run our evaluations. To tune parameters of our classifiers, we used GridSearchCV class in the library. It ran many classifications on the data with different parameters of a classifier and returned the parameters set that give the best performance. For our data set, the best parameters for RF were: n\_estimators=200, max\_features=\emph{log2}, criterion=\emph{entropy}. For SVM classifiers, the best parameters were: C=5, kernel=\emph{rbf}, gamma=\emph{auto}. In all our evaluations, we used ten-fold cross validation, which has been commonly used in~\cite{li2018icare, Frank13, xu2014towards}. The data set was divided into ten equal subsets. In each round, nine subsets were used for training and the remaining subset was used for testing. The overall performance metrics were then averaged. Cross validation is a powerful technique to validate a classifier when the data set is limited.

\subsubsection{Performance of different classifiers}

\begin{figure*}[!t]
    \centering
    \subfloat[AUCs of SVM and RF using \textbf{tap} features on two children groups]{{\includegraphics[width=0.4\paperwidth]{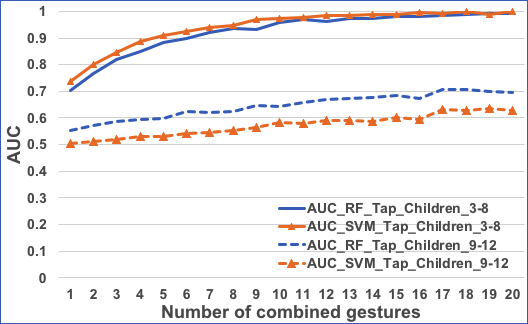} }}%
    \subfloat[EERs of SVM and RF using \textbf{tap} features on two children groups]{{\includegraphics[width=0.4\paperwidth]{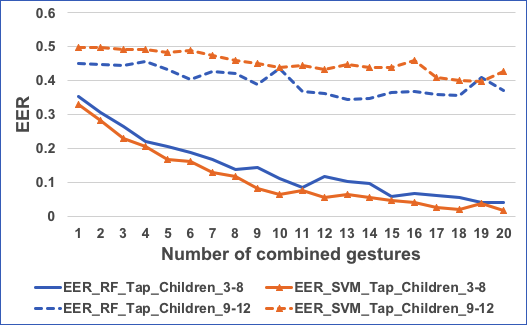} }}%
    \hfill
    \subfloat[AUCs  of SVM and RF using \textbf{stroke} features on two children groups]{{\includegraphics[width=0.4\paperwidth]{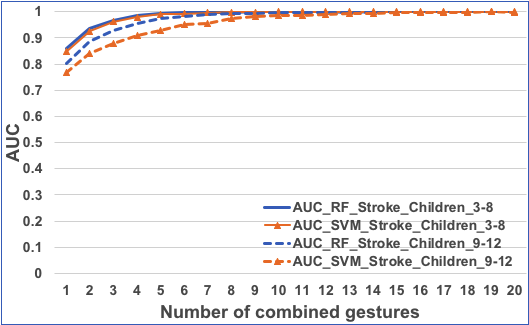} }}%
    \subfloat[EERs  of SVM and RF using \textbf{stroke} features on two children groups]{{\includegraphics[width=0.4\paperwidth]{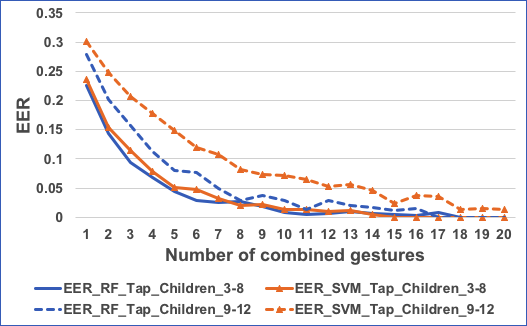} }}%
    \caption{Touch-based approach: Performances of SVM and RF on two children age groups. Using taps or strokes, it's always easier to detect a young child than an older child.}%
    \label{fig:kidsonthephone_performances_different_kid_groups}%
\end{figure*}

\begin{table}[!t]
\centering
\caption{AUC of different classifiers on tap and stroke features}
\label{tbl:kidsonthephone_perf_different_classifiers}
\begin{tabular}{l|l|l}
\textbf{Classifier}         & \textbf{AUC\_Tap} & \textbf{AUC\_Stroke} \\ \hline
Logistic Regression         & 0.60              & 0.73                 \\
Perceptron                  & 0.60              & 0.67                 \\
Multilayer Perceptron (MLP) & 0.63              & 0.79                 \\
SVM (linear kernel)         & 0.63              & 0.75                 \\
\textbf{SVM (rbf kernel)}   & \textbf{0.68}     & \textbf{0.81}        \\
Decision Tree               & 0.60              & 0.73                 \\
\textbf{Random Forest}      & \textbf{0.64}     & \textbf{0.83}       
\end{tabular}
\end{table}
As mentioned before, we ran a grid search for multiple classifiers on aggregated tap and stroke data of both children and adults and show the best performance of each classifier in Table~\ref{tbl:kidsonthephone_perf_different_classifiers}. Here, only one gesture (tap or stroke) was used for a final prediction. As we can see, SVM with an \emph{rbf kernel} and RF provided the best AUCs in both tap and stroke features. Hence, they were chosen to be the classifiers for remaining evaluations.

\begin{figure*}[!t]
    \centering
    \subfloat[AUCs]{{\includegraphics[width=0.4\paperwidth]{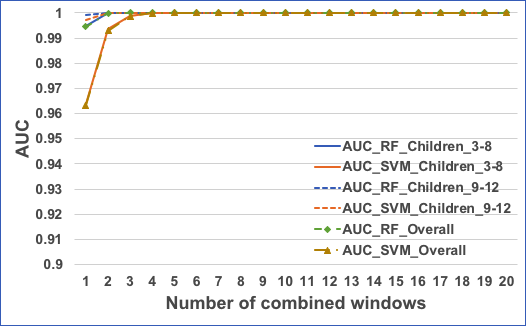} }}%
    \subfloat[EERs]{{\includegraphics[width=0.4\paperwidth]{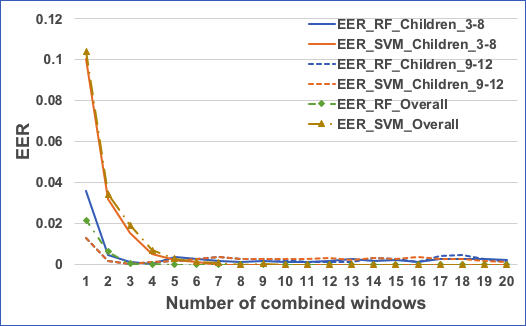} }}%
    \hfill
    \caption{Sensor-based approach: Performances of two classifiers on two children age groups and overall.}%
    \label{fig:kidsonthephone_sensors_performance_kid_groups}%
\end{figure*}

\begin{figure}[!t]
    \centering
    \includegraphics[width=\columnwidth]{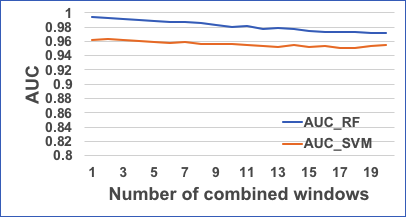}
    \caption{Sensor-based approach: AUCs of two classifiers when features were extracted from different sensor window size. Horizontal axis is the window size in seconds.}
    \label{fig:kidsonthephone_auc_diffent_window_sizes}
\end{figure}

\subsubsection{Classification power of tap and stroke features}
To compare the classification performances when using tap and stroke features we performed ten-fold cross validation evaluations on aggregated tap and stroke data of all children and adults subjects using SVM and RF. The results are presented in Figure~\ref{fig:kidsonthephone_tap_stroke_classification_power}. It is clear that stroke features provide better detection performance for both classifiers. Specifically, with only one stroke, both SVM and RF achieved 0.81--0.83 AUC. In contrast, with only one tap, we can only achieve 0.64--0.68 AUC. In addition, the error rate of both classifiers when using strokes is much smaller than that when using taps: 0.25(RF\_stroke) and 0.26(SVM\_stroke) compared to 0.39(RF\_tap) and 0.37(SVM\_tap). The results are expected and justifiable because strokes provide richer behavioral information than taps.

\subsubsection{Performance evaluation with multiple gestures}
The detection performance can be significantly improved by bundling multiple gestures together to make a final decision. This is clearly demonstrated in Figure~\ref{fig:kidsonthephone_tap_stroke_classification_power}. AUC is considerably improved and EER drastically drops as we bundle more strokes together. AUC reached 0.99 for both classifiers when the number of strokes was 8 (EER\_RF$=0.02$ and EER\_SVM$=0.05$). EER reached 0.002 (or 0.2\%) when the number of strokes was 16. This indicates that by combining a minimum 16 strokes during classification, it is possible to detect children with more than 99\% accuracy while misclassifying adults as children only 0.2\% of the time. Similar effect is observed on tap gestures. However, when combining taps for detection, AUC increased much slower and can only reach 0.9  when the number of taps was 20. In addition, EER was still around 0.1 with 20 taps for a prediction.

\subsubsection{Detection performance using only size feature}
Intuitively, one may guess that it is possible to differentiate a child from an adult by measuring the size of finger tips at the time of performing a gesture. We ran a 10-fold cross validation evaluation on stroke data where only size features were used for classification. The results showed only an AUC of 0.59 (RF) and 0.61 (SVM) compared to 0.83 (RF) and 0.81 (SVM) when using the full feature set proposed in the paper (only one stroke was used for each prediction). The results point out that size is a good feature, but other features are needed for strong classification performance. 

\subsubsection{Detection performance on different children age groups}
The inclusion of a broad range of children' age may harm the classification performance as older children may have significantly different behavior compared to that of younger children. We further divided children subjects into two groups: young children (3--8) and older children (9--12). We then ran evaluation on data of these two groups combined with adults data using 10-fold cross validation. The results are shown in Figure~\ref{fig:kidsonthephone_performances_different_kid_groups}. While Figure~\ref{fig:kidsonthephone_performances_different_kid_groups}(a) and (b) show the AUCs and EERs of SVM and RF on the two children groups when using tap features, (c) and (d) showed AUCs and EERs when using stroke features. As we can see, tap features are not good enough to detect older children. The AUC was about 0.7 while EER was around 0.4 even when we combined 20 taps to make a final decision. Stroke features provided much higher AUCs and lower EERs for both children groups. Specifically, after about 8 strokes, both SVM and RF achieved over 0.99 AUC and less than 0.005 EER for younger children. It took about 14 strokes to achieve this performance with older children. These results are reasonable because younger children possibly are more different from adults in terms of physical (size of fingers and area of fingertip) and behavioral characteristics (swipe faster, make shorter strokes, touch lighter, etc.)

\begin{figure*}[!t]
    \centering
    \subfloat[AUCs]{{\includegraphics[width=0.4\paperwidth]{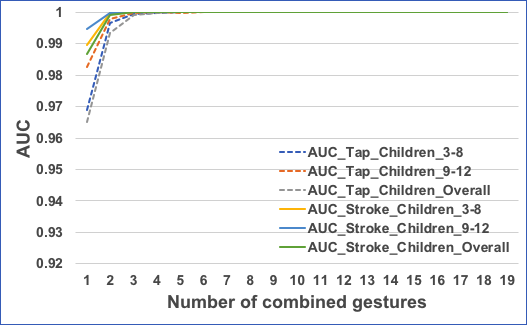} }}%
    \subfloat[EERs]{{\includegraphics[width=0.4\paperwidth]{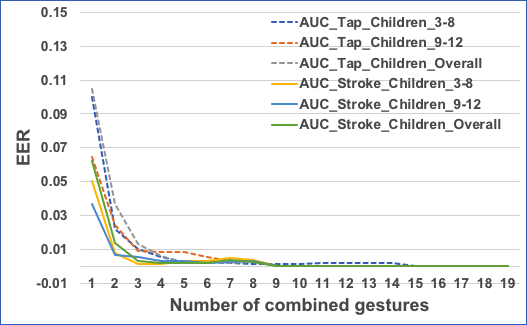} }}%
    \hfill
    \caption{Touch--sensor combination approach: Performances of RF using tap-sensor and stroke-sensor features on two children age groups and overall.}%
    \label{fig:kidsonthephone_combined_features_performances}%
\end{figure*}

\subsection{Sensor-based Classifier}
In this section, we present detailed performance results of the child detection task based on sensor data. We aggregated sensor data into four different groups, namely, a \emph{generic child} group representing all children subjects, a \emph{generic younger child} group representing children of age 3--8, a \emph{generic older child} representing children subjects of age 9--12, and a \emph{generic adult} group containing all the adult subject. Children subjects' data was labeled as positive while adults' data was labeled as negative. Ten-fold cross validation was used in all experiments to evaluate the performance of the classifiers.

\subsubsection{Determine window size}
Since behavioral features are extracted from a time window, it is important to find a good window size. It affects classification performance and also determines the time needed to perform a prediction. We empirically found an appropriate window size as follows. We varied the window from 1 second to 20 seconds. For each window, we extracted features of all subjects and did a ten-fold cross validation using SVM and RF. The results are reported in Figure~\ref{fig:kidsonthephone_auc_diffent_window_sizes}, which suggests that using a window size of 1 second provides us the best AUCs for both classifiers. Thus, we opted to use this window size for the remaining evaluations.

\subsubsection{Detection performance}
We ran ten-fold cross validations on aggregated data sets of each children age group, and of all children with adult subjects. The results are presented in Figure~\ref{fig:kidsonthephone_sensors_performance_kid_groups}. First, as it can be observed that the RF performed better than SVM (lower EERs in all cases). Second, combining multiple windows significantly improves classification performance. Specifically, with only one window, the overall EER of RF and SVM was 0.02 and 0.10 respectively, but it quickly dropped to below 0.002 after combining 5 or more sensor windows for a final detection decision. We also achieved over 0.99 AUC for both classifiers. 

Looking at different children age groups, we can see that both classifiers produced lower EERs for younger children  when the number of combined windows is small (1--5). When more than 5 sensor windows were combined, this difference was negligible. Thus, we conclude that a sensor-based approach is efficient to detect both young and older children.

The results also suggested that a sensor-based approach produced higher performance than a touch-based approach. We can achieve 0.99 AUC and 0.002 EER after 5 seconds with sensor features. But it would take 16 combined strokes to achieve similar performances. In addition, we would never achieve such high performance with taps. There is, however, a drawback of using sensors: it can drain the battery much faster if we continuously track user's movements via sensors. We can mitigate this issue by only recording sensor data whenever the device is unlocked and then periodically checking the user's identity during a session. We do not need to keep polling sensors all the time.

\subsection{Combined Touch--Sensor Classifier}
In this section, we present child detection performance of the combined classifier which is based on both touch and sensor data. Here ten-fold cross validation was also used for all evaluations. The performance obtained is shown in Figure~\ref{fig:kidsonthephone_combined_features_performances}. For readability, we only present results from RF since SVM produced similar figures. As we can see, combining touch and sensor features improved accuracy and reduced error rate compared to those of touch-based features. More specifically, we achieved 0.99 AUC and only 0.01 EER after three consecutive taps and only 0.002 EER after three consecutive strokes. This approach gave us performance as good as using only sensor features, if not better. An important advantage of this approach is that we only need to record sensors when the user performs a gesture on the screen. Thus, it is much less resource intensive compared to the purely sensor-based approach.

\begin{table}[!t]
\centering
\small
\caption{Comparison with previous works using accuracy reported in each paper. AUC is used if accuracy was not reported.}
\label{tbl:kidsonthephone_comparison_with_previous_work}
\begin{tabular}{clll}
\multicolumn{4}{c}{\textbf{Touch-based}}                                                                                                                                                       \\ \hline
\multicolumn{1}{l|}{Age}        & \multicolumn{1}{l|}{}                                & \multicolumn{1}{l|}{Single gesture}                      & Combined gestures               \\ \hline
\multicolumn{1}{c|}{\multirow{4}{*}{3--6}} & \multicolumn{1}{l|}{\cite{vatavu2015child}}        & \multicolumn{1}{l|}{86.5\%, single tap}                  & \textgreater99\%, 7+ taps       \\
\multicolumn{1}{c|}{}                      & \multicolumn{1}{l|}{\cite{hernandez2017detecting}} & \multicolumn{1}{l|}{96\%, single stroke}          & NA                              \\
\multicolumn{1}{c|}{}                      & \multicolumn{1}{l|}{\cite{li2018icare}}      & \multicolumn{1}{l|}{84\%, single swipe}                  & 97\%, 8+ swipes                 \\
\multicolumn{1}{c|}{}                      & \multicolumn{1}{l|}{\textbf{This work}}                       & \multicolumn{1}{l|}{89\%, single stroke}                 & 99\%, 8+ strokes                \\ \hline
\multicolumn{1}{c|}{\multirow{2}{*}{7-12}} & \multicolumn{1}{l|}{\cite{li2018icare}}      & \multicolumn{1}{l|}{0.91 AUC, single swipe}              & 0.97 AUC, 8+ swipes             \\
\multicolumn{1}{c|}{}                      & \multicolumn{1}{l|}{\textbf{This work}}                       & \multicolumn{1}{l|}{0.81 AUC, single stroke}             & 0.99 AUC, 7+ swipes             \\ \hline
\multicolumn{4}{c}{\textbf{Sensor-based}}                                                                                                                                                      \\ \hline
\multicolumn{1}{l|}{}                      & \multicolumn{1}{l|}{}                                & \multicolumn{1}{l|}{Single sensor window}                & Combined windows \\ \hline
\multicolumn{1}{c|}{\multirow{2}{*}{3-12}} & \multicolumn{1}{l|}{\cite{davarci2017age}}   & \multicolumn{1}{l|}{85.3\%, 1 tap window}           & 92.5\% 30 windows           \\
\multicolumn{1}{c|}{}                      & \multicolumn{1}{l|}{\textbf{This work}}                       & \multicolumn{1}{l|}{96\%, 1-second window} & \textgreater99\%, 5 windows    
\end{tabular}
\end{table}

\subsection{Comparison with previous work}
In this section, we compare our results with previous work. 
In terms of performance, Table~\ref{tbl:kidsonthephone_comparison_with_previous_work} shows detailed comparison. Here, we use accuracy as it was used in~\cite{vatavu2015child, hernandez2017detecting, davarci2017age} and AUC was not reported. Accuracy is computed as the ratio between correct predictions over all predictions of testing samples. \cite{li2018icare} used AUC, thus we compared with them using AUC. We used the best figures reported in each paper for comparison. We also recomputed our performance with similar children age groups used in previous work. As we can see, in the touch-based approach, we achieved comparable, if not better than most related work when using a single gesture for detection of children. When combining multiple gestures, we achieved similar accuracy compared to \cite{vatavu2015child} and slightly better than \cite{li2018icare} (for both younger and older children groups). In the sensor-based, our approach achieved much better performance when using a single sensor window as well as a combination of multiple ones compared to ~\cite{davarci2017age}. These results confirm the effectiveness of the proposed approaches.

\section{Discussion and future work}
\label{sec:kidsonthephone_discussions}

We have presented three different behavior based children detection methods on smart devices, namely, touch-based, sensor-based and a combination of sensor and touch-based approach. The touch-based approach leverages tap and stroke patterns recorded from a user's touch interactions on a mobile device. The sensor-based approach tracks a users' hand movement from a stream of data recorded by motion sensors while the user is  handling the device. The combination approach utilizes both touch-based and sensor-based features to improve detection accuracy. To this end, we acquired a new data set of natural touch behavior of 50 children and adults subjects. The main findings of this work are as follows:
\begin{itemize}
\item The work establishes the fact that it is indeed possible to automatically detect presence of children on mobile devices with high accuracy. The advantages of the proposed techniques are as follows: first, they are implicit in nature, thus do not need active participation of the subjects. Second, the techniques are robust as multiple gestures, multiple time windows and multiple data sources are used to make the final decision. 
\vspace{-0.5em}
\item The results show good performance on the child detection task with over 0.99 AUC and less than 0.5\% EER for all the three approaches investigated (after 8 consecutive strokes, or after 5 sensor windows of size one second, or after 3 consecutive strokes in the combination approach).
\vspace{-0.5em}
\item Comprehensive evaluation results with different age group of children indicated that not only the younger children (3--8 years) but also the older ones (9--12 years) can be classified efficiently with the proposed classification frameworks.
\end{itemize}

While focus of the current work is on developing techniques to distinguish between children and adults based on touch behavior, not on how the technique will be used, we briefly discuss broader implications of this work in the next section.
 
\subsection{Broader implication of this work}

The proposed technique can be employed in designing various applications like methods for children online safety, parental control, user interface (UI) design for early education, etc. If this technique is to be used with parental control, it must be integrated carefully considering the multi-faceted nature of the issue. Various reports on child development have pointed out that, at very young age, different forms of parental mediation are beneficial to healthy development~\cite{TVViewingTime, HealthyChildren}. There are five parental mediation strategies: (1) restrictive mediation~\cite{mendoza2009surveying}, (2) co-use~\cite{nikken2014developing}, (3) active mediation~\cite{padilla2012parents}, (4)deference~\cite{padilla2012parents}, and (5) supervision~\cite{nikken2014developing}.  It is evident that the opportunities for parental controls will unfold differently for various age groups and in various contexts. However, all of the common types of mediation are recommended by the child development research community~\cite{Strasburger1012}, but restrictive mediation in particular is strongly encouraged for very young children (though their stance on this subject is also in flux, see~\cite{brown2015beyond}). However, restrictive measures come with several drawbacks. The one-sided focus on protection may even be detrimental to children rights and well-being~\cite{janssens2015parents}. From this viewpoint, children may not benefit from constant parental protection~\cite{hartikainen2016should}. 

Current research has emphasized the importance of balancing the needs of parents to monitor and manage usage with the needs of children to maintain the agency and autonomy that children need to develop into self-dependent adults~\cite{hartikainen2016should, mazmanian2017okay, zaman2016parental}. Several studies have reported that appropriate balance in parental mediation can mitigate the negative aspects of digital media~\cite{nouwen2015value, ko2015familync, wisniewski2015resilience, cantor2003media, nathanson2000reducing} and help to develop digital literacy and practical abilities in the use of digital devices~\cite{austin1993exploring}. It has been shown that mediation enforced by technology are more effective than mediation enforced by parents~\cite{hiniker2016screen}. From this view point, it is beneficial if the device can automatically detect interaction from children, which is focus of this work.


\subsection{Limitations}
\label{subsec:kidsonthephone_limitation}

Although impressive performance of the proposed approach indicates inherent strength of the behavior-based model, there still remain several limitations  as identified below.

First, although we have acquired a relatively bigger and more diverse data set than previous work, it is still limited, and may not be sufficient to generalize the model of behavior for children and adults. Nevertheless, we have shown the potential of our approaches. We believe we can train a more robust model with more data, for which we plan to conduct in future work.

Second, this study has not considered the case of subjects with physical impairments or older individuals. Older individuals often have lower motor precision similar to children. It may affect the performance of the classifier substantially, as touch gestures are not only characterized by the dynamics but also by other physical properties, like pressure applied on screen and area of touch. We envision our technique being used in situations where there potentially are children users and would not be activated on a phone that has adult users with physical impairments. This class of users are a common issue for touch interaction based studies. 

Third, we have not trained and evaluated our models in different contexts. For example, when users are walking, their touch and sensor readings will be affected, which may lead to differences in their behavior model. This is also a common problem in touch and sensor-based user authentication works. We can use activity recognition techniques~\cite{su2014activity} or contexts detection~\cite{hayashi2013casa} to determine current user activity and only perform classification  in a static mode. Another approach is to take into account the variance in the training data in different contexts and activities and train different models correspondingly. 

In addition, current work has not examined other types of gestures like multi-finger gestures and the possibility of fusing different classifiers of different gestures for better and faster detection. Last but not least, we have not evaluated our models across different devices and different vendors. For example, how would a model built on mobile phones perform on a tablet? Hernandez et. al~\cite{hernandez2017detecting} suggested this might work, however, a more thorough analysis on a bigger data set is needed. 

\balance{}
\bibliographystyle{SIGCHI-Reference-Format}

\end{document}